\documentclass[notitlepage,aps,prd,amsmath,floats,floatfix,
twocolumn,
superscriptaddress,nofootinbib,showpacs]{revtex4-1}

\usepackage{amsfonts,amsmath,units,wasysym,epsfig,graphicx,verbatim,color,subfigure,graphicx}
\usepackage[export]{adjustbox}
\usepackage{amsmath}
\usepackage{amssymb}
\usepackage{amsfonts}
\usepackage{float}
\usepackage{bm}
\usepackage{color}
\usepackage[normalem]{ulem}
\graphicspath{{Images/}}

\usepackage{natbib}
\usepackage{color,soul}
\usepackage{hyperref}
\usepackage{marginnote}
\usepackage{braket}
\usepackage[none]{hyphenat}
\usepackage{bm}
\usepackage{comment}
\usepackage{multirow}
\usepackage{array}

\usepackage{colortbl}
\usepackage{tabulary}

\newcommand{\SU}{\affiliation{Department of Physics, Syracuse University, Syracuse, New York 13244, USA}}

\begin{document}

\title{High Frame-Rate Phase Camera for High-Resolution Wavefront Sensing in Gravitational-Wave Detectors}
\author{Erik~Mu\~niz}\SU
\author{Varun~Srivastava}\SU
\author{Subham~Vidyant}\SU
\author{Stefan~W.~Ballmer}\SU

\date{\today}

\begin{abstract}
We present a novel way of wavefront sensing using a commercially available, continuous-wave time-of-flight camera with QVGA-resolution.
This CMOS phase camera is capable of sensing externally modulated light sources with frequencies up to 100 MHz. The high-spatial-resolution of the sensor, combined with our integrated control electronics, allows the camera to image power modulation index as low as -62~dBc/second/pixel. The phase camera is applicable to problems where alignment and mode-mismatch sensing is needed and suited for diagnostic and control applications in gravitational-wave detectors. Specifically, we explore the use of the phase camera in sensing the beat signals due to thermal distortions from point-like heat absorbers on the test masses in the Advanced LIGO detectors. The camera is capable of sensing optical path distortions greater than about two nanometers in the Advanced LIGO input mirrors, limited by the phase resolution. In homodyne readout, the performance can reach up to 0.1 nm, limited by the modulation amplitude sensitivity.
\end{abstract}
\keywords{Phase Camera, Wavefront Sensing, High-Resolution Sensors}
\maketitle

\section{Introduction}\label{sec:intro}

The second generation of gravitational-wave detectors Advanced LIGO~\cite{aLIGO2015} and Advanced VIRGO~\cite{Acernese_2014} have been observing the compact binary mergers for over five years. The first direct detection of gravitational waves from the inspiral and merger of two binary black-holes happened on September 14, 2015, opening up the field of gravitational wave astronomy~\cite{GW150914}. Since then there have been two sets of upgrades to improve the sensitivity of the Advanced LIGO and VIRGO detectors~\cite{comm_O3} and the three observational science runs have confirmed over 50 gravitational wave signals from binary collisions~\cite{abbott2019gwtc, abbott2020gwtc}.

The Advanced LIGO and VIRGO detectors are dual-recycled Fabry-Perot Michelson interferometers capable of measuring relative peak displacement between the two interferometer arms to less than one attometer~\cite{aLIGO2015, Acernese_2014}. To enable this precise measurement, the core and auxiliary optics are seismically isolated and the feedback loops are set up to actuate on translational and angular degrees of freedom~\cite{aLIGO2015}. The error signals to control the optics are derived using Pound-Drever-Hall (PDH) technique~\cite{black2001introduction, fritschel1998alignment} from 9~MHz and 45~MHz sidebands modulating the carrier laser beam.
These error signals are acquired using radio-frequency (RF) photodiodes - single-segment diodes for length sensing, and four-segment quadrant photodiodes for angular sensing. Similarly, beam waist position and size sensing can be done using bulls eye segment diodes \cite{mueller2000determination}, mode converters \cite{magana2019sensing} and aperture-based schemes \cite{miller2014length}. Alternatives to this readout scheme using jitter modulation techniques have also been proposed \cite{Fulda:17}.

Since any degradation in the optical wave front will degrade these signals, significant effort has gone into ways to image the RF optical beat pattern of the readout beam for diagnostic purposes. Previously developed sensors for the LIGO and Virgo detectors include scanning-type \cite{Goda:04, van2016advanced, agatsuma2019high} and optical lock-in \cite{Cao:20, brown2021differential} phase cameras, all of which require additional optical elements ahead of the photo sensor.

In our work, we redesign the traditional time-of-flight camera to directly record these amplitude-modulated laser beat signals in QVGA resolution (320 $\times$ 240 pixels). We achieve this by phase locking the OPT8241 time-of-flight camera to the external reference oscillator that modulates the laser.  This approach has the benefit of requiring minimal optical components and beam shaping,  
making the CMOS phase camera an excellent diagnostic tool for gravitational-wave detectors that can be easily deployed at any optical port of the interferometer. We also demonstrate that this phase camera can be used for generic wavefront sensing.

The sensitivity of interferometric gravitational wave detectors such as Advanced LIGO scales with the amount of power circulating in the arm cavities. Following the O3 upgrade, the Advanced LIGO detectors reached 200~kW of power in the arm cavities~\cite{comm_O3} for the first time. The high power in the arm cavities exposes point defects in the coatings of the test masses. Under high power these point defects, or point absorbers, burn into the coating of the Advanced LIGO test masses~\cite{brooks2021point}, locally heating and deforming the optic. This has proven to be problematic, limiting power buildup in the arm cavities and dark port contrast, both of which limits the sensitivity of the Advanced LIGO detectors. Specifically, the thermo-refractive and thermo-elastic surface deformations due to the point absorbers induce optical path distortions affecting the phasing of the carrier and the modulating sidebands~\cite{brooks2021point}. For the carrier, this detuning results in higher-order spatial modes being resonant in the arm cavity, causing a loss in the detector sensitivity. Additionally, the non-resonant RF sidebands couple into higher-order spatial modes and degrade the alignment and control error signals error.

We aim to use the CMOS phase camera to image the deformations in the laser beat signals that arise from these point defects. In the paper, we briefly describe the principle of operation of time-of-flight cameras in section~\S\ref{sec:ppl_ToF}. Next, in section~\S\ref{sec:electronics} we describe the hardware and software changes made to reconfigure the time-of-flight camera into a CMOS phase camera. The experimental layout to test the sensitivity of the phase camera and the corresponding noise model to measure amplitude modulated signals is described in section~\S\ref{sec:exp_sense}. We model the point absorber to estimate the change in beat signal and correlate it with the sensitivity of the camera in section~\S\ref{sec:pointAbs}. Lastly, we discuss the applicability of the CMOS phase camera in Advanced LIGO and A+ detectors in section~\S\ref{sec:results}.

\section{Principle of Operation of time-of-flight cameras}\label{sec:ppl_ToF}

\begin{figure}[htp]%
\center
\includegraphics[width=0.9\linewidth, frame]{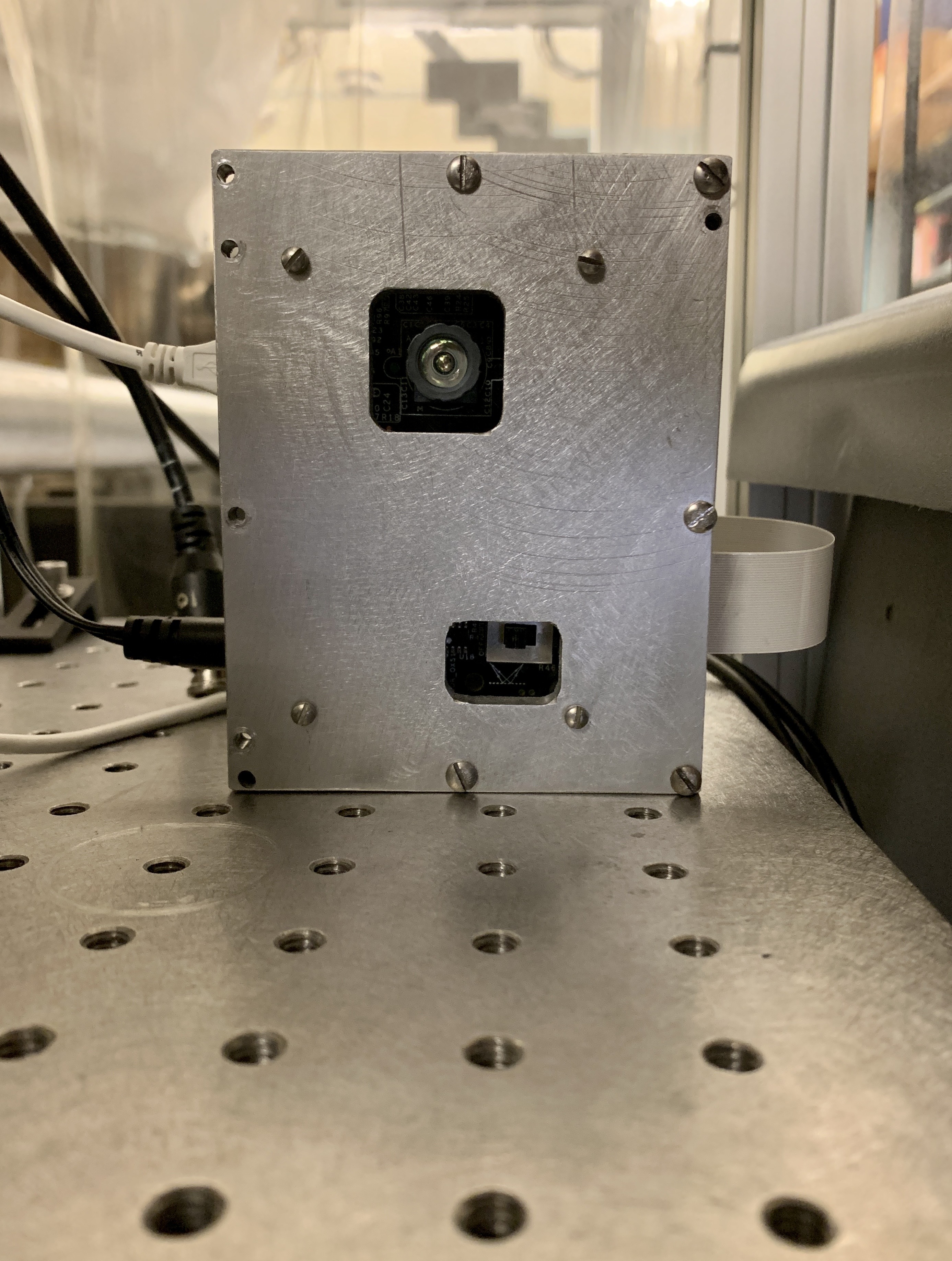}
\caption{The prototype CMOS phase camera based on the commercially-available OPT8241 time-of-flight sensor. The custom-built enclosure ensures shielding from interference from unwanted RF signals.}
\label{fig:pcdemo}
\end{figure}

Time-of-flight cameras employ a 2D array of pixels with depth sensing capabilities to independently measure the distance of a particular object in the field of view. Commercially available time-of-flight cameras typically have integrated illumination sources. They rely on either pulse-modulation, directly measuring the pulse travel time, or on continuous-wave amplitude-modulation, measuring a phase delay via demodulation with a local reference. The measured phase delay provides depth information. Since we intend to use the sensor to record the beat map of laser fields we selected the second type sensor for our research.

The amplitude $A$, phase $\phi$ and DC value $I_0$ of the amplitude modulated beam can be found by measuring the gated photo current $C$ at four demodulation phases ($\Phi$ =  0, $\pi/2$ , $\pi$ , $3\pi/2$)~\cite{lange:00}. We have
\begin{equation}\label{eqn:amp_rf}
    \text{A}= \frac{\sqrt{\{C(0) - C(\pi)\}^2 + \{C(\pi/2) - C(3\pi/2)\}^2 }}{2}
\end{equation}
\begin{equation}\label{eqn:phase}
    \phi= \measuredangle [(C(0) - C(\pi))~+i(C(\pi/2) - C(3\pi/2))]
\end{equation}
\begin{equation}\label{eqn:amp_dc}
    I_0=\frac{C(0) + C(\pi/2) + C(\pi) + C(3\pi/2)}{4}
\end{equation}

\section{Motivation and Design}\label{sec:electronics}

For application as a wavefront sensor in gravitational-wave interferometers the time-of-flight camera needs to accept the interferometer RF local oscillator signal as an external reference oscillator source for demodulation.
In continuous-wave amplitude-modulated time-of-flight cameras, both the illumination source and pixels are typically driven using the same internal, radio-frequency (RF) oscillator source. \citet{Shrestha:2016} show that the time-of-flight camera OPT8221 offers flexibility for operation as a standalone demodulation camera with an external signal driving the illumination source and the demodulation within the camera. 

The design proposed here uses the OPT8221 camera evaluation board (OPT8241-CDK-EVM), which consists of two primary components: the sensor (OPT8241) and a programmable controller (OPT9221). The OPT9221 is a companion chip to the OPT8241 and is responsible for setting register functions and processing raw data. The sensor is a standard 320 $\times$ 240 pixel array (QVGA format) capable of operating at frame rates up to 150 frames per second (fps), although live-streaming with Voxelviewer software~\cite{voxelviewer} is limited to 60 fps. Neighboring pixels are separated by 15~$\mu$m and are capable of demodulation up to 100 MHz.
In each pixel the charge carriers are sorted into two separate charge storage wells, depending on the state of the local oscillator \cite{opt8241sysdesignguide,schwarte1997new}. The two wells are simultaneously read out and their difference is digitized, removing the DC component from the RF readout. Charge separation becomes less efficient at higher frequencies \cite{LangeThesis}, leading to a reduction in the demodulation amplitude at higher demodulation frequencies.  

The camera board can be programmed to provide 12-bit amplitude and phase, as well as 4-bit ambient values in QVGA format. The OPT8241 also features a 850 nm NIR band-pass filter covering, which has a small transmission of about 5\% at 1064 nm for normal incidence~\cite{opt8241datasheet}.  We use the sensor at 1064 nm, the laser wavelength used in the LIGO, VIRGO, and KAGRA gravitational wave detectors~\cite{aLIGO2015, Acernese_2014, aso2013interferometer}.

For application as an active wavefront sensor the camera board must be reconfigured. First, the internal modulation block of the camera board is disabled via the appropriate configuration of register settings in the OPT9221, allowing it to accept an external reference signal. This reference signal needs to be appropriately stepped in quadrature phase, with $\Phi = 0^\text{o}, 90^\text{o}, 180^\text{o}, 270^\text{o}$. The frequency of the reference signal can also be dynamically changed between video frames, permitting imaging beat signals from multiple sidebands simultaneously.
Thus, the CMOS phase camera is capable of measuring both the spatial and temporal characteristics of the illuminating beam in real-time.

\begin{figure}
    \centering
    \includegraphics[width=\linewidth]{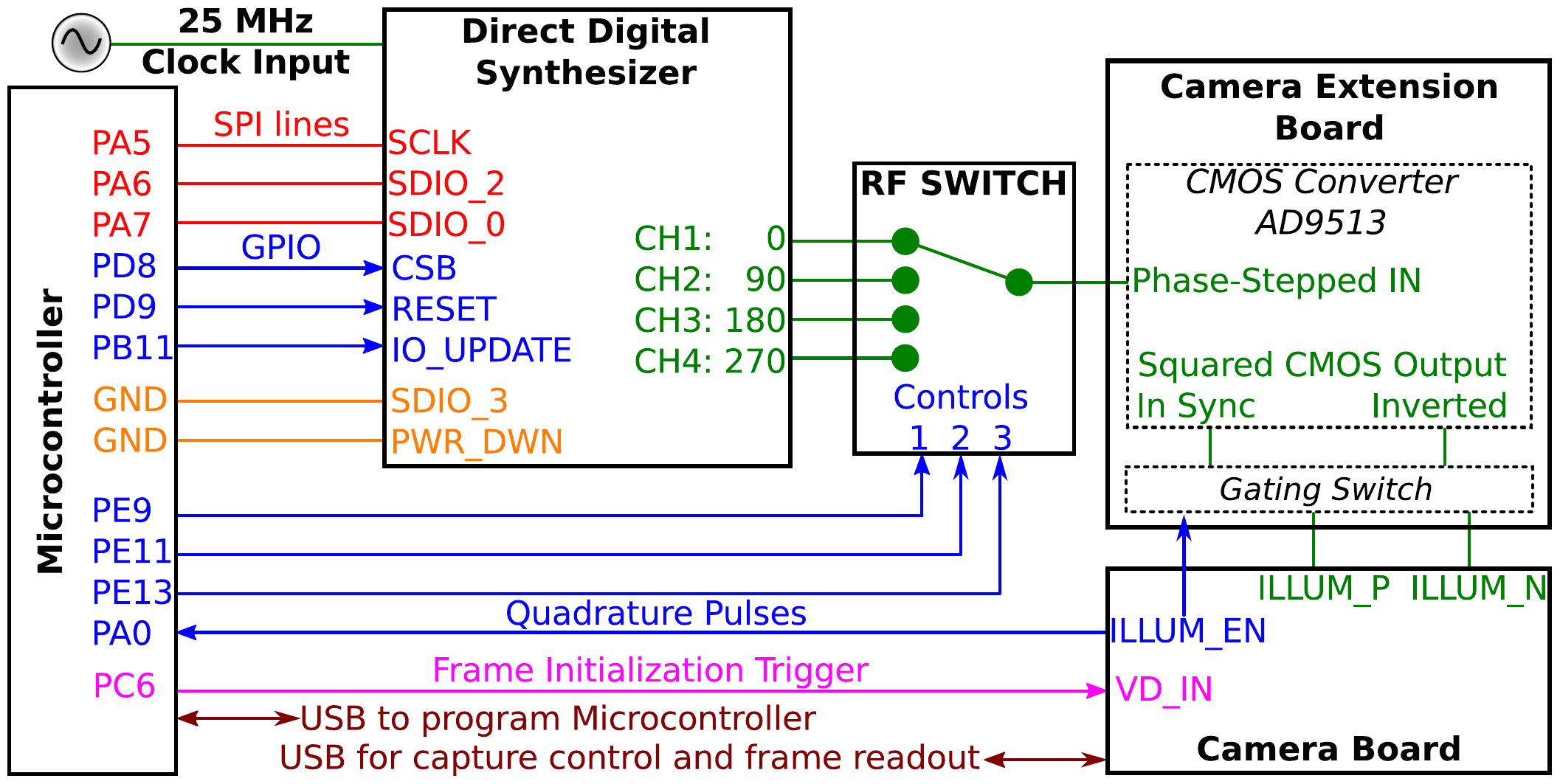}
    \caption{An overview of the phase camera signal chain and design components. The microcontroller acts as the host device controlling waveform generation and image frame capture. The microcontroller initializes the direct digital synthesizer board. Quadrature phase-stepped signals are generated on the four analog output channels of the direct digital synthesizer board. These signals are sent to an RF switch, which passes the appropriate signal via CMOS converter and gating switch to the camera. This gating switch removes the RF signal during the sensor readout to reduce the electronics noise. 
    Every frame is initiated by a frame initialization trigger sent by the microcontroller to the camera board. The camera responds by sending a series of quadrature pulses back to the microcontroller, which are used in an interrupt sequence to set the control signals for the RF switch. }
    \label{fig:signalchain}
\end{figure}

In this section, we present the schematic design for the CMOS phase camera. The various electronic components, along with their functionality, are listed below and illustrated in Fig.~\ref{fig:signalchain}:
\begin{itemize}
    \item {\it Microcontroller}: We use the ARM Cortex-M4-based STM32F407 as a master device. It is programmed to control waveform generation on the Direct Digital Synthesis (DDS) board. The microcontroller transmits a data sequence to initialize each of the individual channels on the DDS board via SPI protocol.
    The microcontroller also initiates the frame capture sequence by sending an initial trigger pulse to the camera board. Depending on the register settings of the OPT9221 controller, the camera board responds by sending a series of quadrature (quad) exposure pulses, which trigger an interrupt sequence on the microcontroller. This interrupt sequence is used to set the logic levels on the RF switch and/or the synchronization flip-flop to output the appropriately phase-stepped signal corresponding to the quad exposure pulse.
    
    \item {\it Direct Digital Synthesizer (DDS) board}: Waveform generation is handled by the Analog Devices Direct Digital Synthesizer board (AD9959). The AD9959 has four independent  synchronized output channels with tunable amplitude, frequency and phase. The external reference local oscillator used to modulate the interferometer input laser serves as the reference clock signal to the DDS board for synchronization.
    The microcontroller sets frequency, phase and amplitude registers on the DDS board for each channel and activates the settings via a separate I/O-update line.
    For normal operation the four channels are set to a phase of $\Phi =0^\text{o},~90^\text{o},~180^\text{o},~270^\text{o}$ with respect to the reference clock at the desired demodulation frequency during initialization.
    
    \item {\it RF switch}: A single-pole, four throw RF switch from Mini-Circuits (JSW4-272DR+) is used to switch between the four phased channels from the DDS board, selecting the camera demodulation phase. The RF switch is controlled by the microcontroller. The DDS board and RF switch together can provide an externally referenced and phase-stepped oscillator between 5~MHz to 500~MHz.

    \item {\it Synchronization Flip-Flop}: The DDS board register update is only synchronized to the internal reference clock (SYNC\_CLK), which is running at a higher frequency than the external reference signal. To avoid random phase jumps on update, a flip-flop is used to trigger the update of the DDS registers synchronized with the rising edge of the external reference signal. There are two schemes to operate the phase camera.
    First, the four channels of the DDS board are set up during camera initialization, and the flip-flop synchronization to the reference oscillator guarantees the same relative phase on every camera startup. The RF switch is then used to select the channel with the desired demodulation phase. Since this method does not rely on a synchronized register update during operations it is more robust, but also limited to four demodulation phases and a single frequency.
    In the second approach, we use the flip-flop to actively update the registers and phase-step the output of one of the DDS channels in accordance with the quad pulses from the camera. This scheme, discussed in-depth in section~\S\ref{appsec:ff_scheme}, allows the camera to image the amplitude-modulated beat at different frequencies in real-time, and permits camera operation with six quads, reducing potential cross-talk from harmonics of the modulation frequencies.
    
    \item {\it CMOS Converter}: To convert the sinusoidal reference signal to a CMOS logic level clock, i.e.~`squaring the clock', we use a high-speed clock distribution integrated circuit (Analog Devices AD9513).
    
    \item {\it Gating Switch}: We use an additional high-speed buffer chip (Texas Instrument's SN74LVC126A) in the signal chain to gate the output demodulation waveform during each quad exposure. Turning the RF signal off during the readout phase between camera exposures significantly reduces the camera electronics noise.
\end{itemize}
All firmware was developed within the Keil $\mu$Vision IDE and flashed to the microcontroller via a USB interface. This allows the microcontroller to run as the master device after an initial reset event. 

\section{Performance Characterization}

\subsection{Experimental Setup}\label{sec:exp_sense}

\begin{figure}
\center
\includegraphics[width=\linewidth]{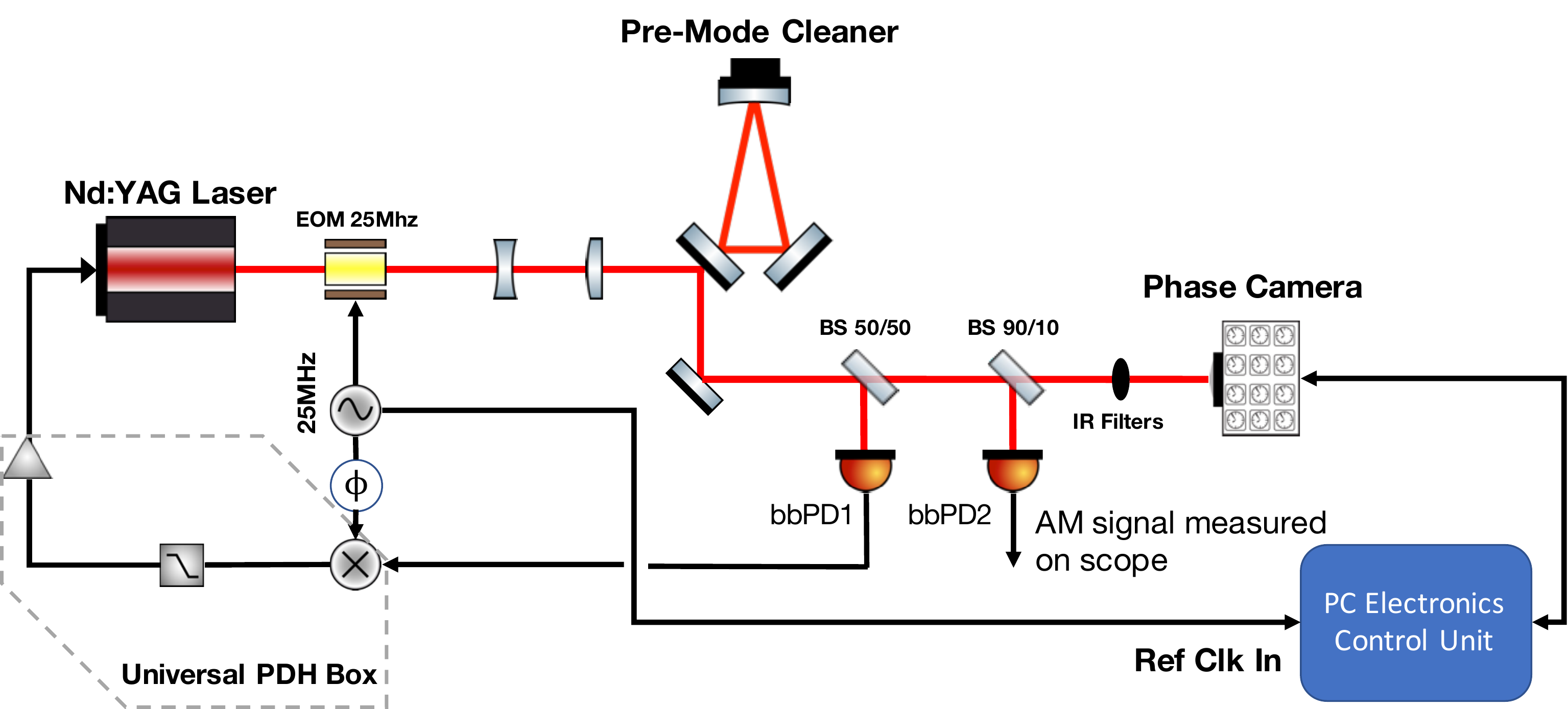}
\caption{The experimental layout used for testing and quantifying the noise levels of the phase camera. The laser is stabilized by locking to the pre-mode cleaner cavity using the PDH technique. The amplitude-modulated beat signal is generated in the reflection of the cavity when it is locked to one of the 25~MHz sidebands. The bbPD1 is the photodiode sensor used for PDH locking. The bbPD2 is used in calibration and measures the power modulation index of the beam incident on the camera. The total power incident on the camera is attenuated to $\sim$10$~\mu$W.}
\label{fig:exp_setup}
\end{figure}

To characterize the performance of the CMOS phase camera we used the experimental setup shown in Fig~\ref{fig:exp_setup}. 
A reference triangular cavity, previously used as 
pre-mode cleaner cavity (PMC) in the initial LIGO interferometers until 2010, is used to stabilize the frequency of a 1064 nm Nd:YAG laser via Pound-Drever-Hall (PDH) locking. The cavity has a finesse of 165 and features a piezoelectric-actuated mirror at the apex which allows for tuning of the cavity resonance. The carrier field, $E_{c}=E_{0}e^{i\omega_{c}t}$, is phase-modulated using a resonant electro-optic modulator referenced to a 25-MHz local oscillator.  This adds sidebands to the optical carrier field offset at the modulation frequency $\Omega/2\pi$. The peak-to-peak drive voltage of the local oscillator determines the amplitude of the phase modulation index $\Gamma$. The field incident on the cavity is given by
\begin{equation}
    E_{in} = E_{c}e^{i\Gamma \cos\Omega t} \approx E_{c} \left(1 + \frac{i \Gamma}{2} e^{i \Omega t} + \frac{i \Gamma}{2} e^{-i \Omega t} \right)
\end{equation}
To produce a significant amount of amplitude beat signal between carrier and sideband the cavity length was tuned to resonate on a single sideband. Under the simplifying assumption that one sideband passes the cavity $100\%$, while the other light gets reflected $100\%$, the reflected field becomes to first order
\begin{equation}
    E_{ref} \approx E_{c} \left(1  + \frac{i \Gamma}{2} e^{-i \Omega t} \right) ,
\end{equation}
which is then passed through a 50/50 beam splitter where half of the laser light is reflected onto a single-element broadband RF phototdiode (bbPD1) for PDH locking and the other half is directed to the camera. From here, the beam power is split again with 90\% of the light going to the camera and 10\% incident on a second photodiode (bbPD2) used for calibration -- discussed more in section~\S\ref{subsec:calibration}.  The power incident on the camera is thus amplitude modulated with power modulation index $\Gamma$:
\begin{align}\label{eqn:powermodulation}
    \nonumber
    I_{PC} &= |E_{ref}|^{2} \\
           &\approx |E_{c}|^{2}\big(1 + \Gamma_{} \sin\Omega t\big) 
\end{align}
Frequency terms greater than the modulation frequency are filtered by the bandwidth of the camera. 
In this single-sideband scenario power modulation index $\Gamma$ is also related to the sideband-to-carrier ratio ${\rm SCR}$ via
\begin{equation}\label{eqn:dBc}
    {\rm SCR \, [dBc]} = 20 \log_{10} \left( \frac{\Gamma}{2} \right), 
\end{equation}
where we use the IEEE's definition \cite{807679}.
Unlike typical photodetectors, the measured photocurrent is demodulated at $\Omega$ in each pixel on the sensor array. 
The corresponding amplitude and phase maps are then constructed using the measured values of I and Q for each frame. The CMOS phase camera in our experiment was set to capture at a frame rate of 7~Hz with total exposure time of 32~ms, although frame rates of up to 60~Hz are supported.

\subsection{Calibration}\label{subsec:calibration}

\begin{figure}
\center
\includegraphics[width=\linewidth]{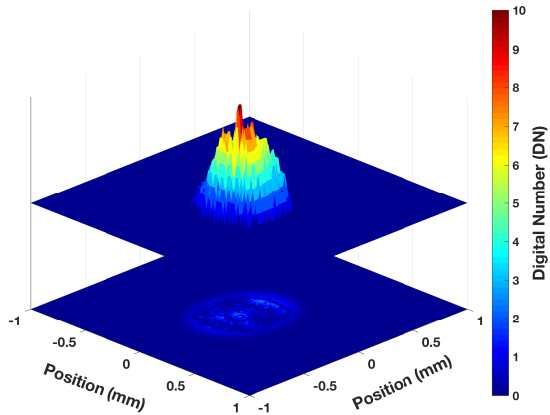}
\caption{The DC intensity profile, which is used in calibration, is plotted as a function of sensor position. The residual of the Gaussian fit to the DC profile is shown in the plane on the bottom. The dominant 4-bit digitization noise in the DC output is noticeable. In contrast, the RF output is digitized with 12-bit resolution.}
\label{fig:dcfit}
\end{figure}

For each frame of the 320 $\times$ 240-pixel array, the respective amplitude, phase, and ambient readout channels of the camera are expressed in terms of counts of the analog-to-digital converter or digital numbers (DN). We perform a calibration procedure to determine a calibration factor, $\kappa$, which relates the total number of generated photoelectrons in an image to the corresponding DN value. The calibration procedure is as follows:
\begin{enumerate}
    \item We use the broadband photodiode (bbPD2, see Fig.~\ref{fig:exp_setup}), to determine the modulation index of the beam incident on the camera. We measured its DC transimpedance $Z_{dc}$ to be 1975~$\Omega$ and its RF transimpedance $Z_{rf}$ at 25~MHz, the frequency of the local oscillator, to be 4750~$\Omega$. Then, the power modulation index ($\Gamma_{m}$) of an arbitrary beam incident on \textit{bbPD2} can be calculated via
    \begin{equation}\label{eq:mod_index}
        \Gamma_m = \frac{1}{2} \cdot \frac{V_{pp}^{rf} \cdot Z_{dc}}{ Z_{rf} \cdot V_{dc} }
    \end{equation}
where the peak-to-peak RF signal ($V_{pp}^{rf}$) and the mean DC signal ($V_{dc}$) measured by \textit{bbPD2} are read on an oscilloscope. The factor of 1/2 converts peak-to-peak to amplitude.
    \item To quantify the number of photons incident on the camera we measure the incident power ($P_{in}$) using a calibrated Thorlabs power meter. Since the NIR filter on the camera has a 5$\%$ transmission at 1064~nm at normal incidence~\cite{opt8241datasheet}, the total power incident on the camera after the filter ($P_{in}^{sensor}$) can be calculated. The ambient channel of the camera measures the profile of the incident beam. We fit a Gaussian profile normalized to the $P_{in}^{sensor}$ to get the Gaussian beam parameters and calculate the power density of the beam incident on the sensor, $p_{in}^{sensor}(x,y)$, see Fig.~\ref{fig:dcfit}. The number of photons per area incident on the sensor during the exposure time $T_{exp}$ can then be estimated by
    \begin{equation}\label{eq:no_photons}
        n_{p} (x,y) = \frac{p_{in}^{sensor} (x,y) \cdot T_{exp}}{h\nu}
    \end{equation}
    The exposure time is the sum of each of the quadrature exposure time during which the signal is integrated. 
    \item The number of photoelectrons generated in each pixel is related to the number of photons via the quantum efficiency $\eta$, which for the OPT8241 sensor is approximately 2$\%$ for 1064 nm light ~\cite{LangeThesis}. Thus we have
    \begin{equation}\label{eq:no_electrons}
        N_e = \int_{A_{pixel}} \!\!\!\!\!\!\!\!\!\!\!\! n_{e} (x,y)~dx dy = \int_{A_{pixel}} \!\!\!\!\!\!\!\!\!\!\!\! \eta \cdot n_{p} (x,y)~dx dy
    \end{equation}
    \item Finally, the calibration factor $\kappa$ is given by the ratio of the total number of photoelectrons generated to the sum of digital numbers reported in a region of interest (ROI):
    \begin{equation}\label{eq:calib}
        \kappa_{DC} (e^{-}/DN) = \frac{ \Sigma_{ROI} (N_{e}) } {\Sigma_{ROI} (DN_{DC})}
    \end{equation}
    Similarly, the AC calibration is given by
    \begin{equation}\label{eq:calib}
        \kappa_{AC} (e^{-}/DN) = \Gamma_m \frac{ \Sigma_{ROI} (N_{e}) } {\Sigma_{ROI} (DN_{AC})},
    \end{equation}
    where $\Gamma_m$ is the modulation index from Eq.~\ref{eq:mod_index} and we divide by the camera output in the AC readout, $DN_{AC}$.
    We estimate a calibration factor of $\kappa_{DC}$ of $1.9\cdot10^5~e^{-}/DN$  and $\kappa_{AC}$ of $1.0\cdot10^{3}~e^{-}/DN$.

\end{enumerate} 

\subsection{Noise Sources}\label{subsec:noise}
The CMOS phase camera offers both the AC and DC readout channels, which are susceptible to noise sources typical to CMOS image sensors. These noise sources can either be temporal or spatial. Temporal noise arises due to the electronic noise from resistive components (i.e. pixel reset noise), dark noise, photon shot noise and other electronics noise related to the image sensor. Spatial noise is primarily due to the pixel-to-pixel imperfections, that is gain and threshold variations in the sensor array. 

For the DC readout, the 4-bit digitization of the ambient channel results in large analog-to-digital rounding errors. As a result, the ADC quantization noise is the limiting source of noise for the DC readout. In contrast, the AC readout provides a 12-bit resolution, sufficient for resolving small fluctuations of the measured signal. For the AC readout, the different noise sources of the phase camera are presented below:

\begin{enumerate}
    \item \textit{Shot Noise} in the camera translates to an effective fluctuation in the measured number of photoelectrons generated in the AC readout channel. 
    The shot noise scales as the square root of the number of electrons generated during the exposure time. The AC shot noise can be estimated as
    \begin{equation}
        \sigma_{SN}~(e^-) \approx \sqrt{2 N_e} 
    \end{equation}
    The factor of $\sqrt{2}$ is due to the demodulation. Shot noise is a fundamental limitation, but our CMOS phase camera is not shot noise limited.
    \item \textit{Electronic Noise} includes all noise sources involved in charge conversion and signal processing within the image sensor. Two prominent noise sources in the CMOS-based sensors are the amplifier noise and the pixel reset noise. Both arise due to the Johnson noise associated with the reset transistor in each pixel. Therefore, the electronic noise (in DN) can be estimated from non-illuminated regions in the frames, using the temporal standard deviation across multiple frames. 
    The effective electronic noise can be reduced by frame- and/or pixel-averaging (i.e. temporal or spatial averaging).
    \item \textit{Fixed Pattern Noise} (FPN) describes the spatial noise associated with non-homogeneity between neighboring pixels on the sensor. Fixed pattern noise is dependent on the signal on the sensor and for a given signal can be expressed as~\cite{georgiev2015fixed}
    \begin{equation}
        \sigma_{FPN} = \alpha~\Tilde{C} + \beta
    \end{equation}
    where $\alpha$ is the gain and $\beta$ is a column or row offset in the readout. $\Tilde{C}$ represents the true DC signal with additive Gaussian noise, which represents the temporal fluctuations of the signal contributing to the fixed pattern noise due to the gain. Under uniform illumination, one can estimate the pixel fixed pattern noise (in DN) as the spatial standard deviation from the mean. Using the calibration factor calculated in Section~\S\ref{subsec:calibration}, we can express the fixed pattern noise in terms of the number of electrons. As $\sigma_{FPN}$ varies as a function of the signal intensity, trivial background frame subtraction will not work to improve the SNR. Spatial averaging of frames can reduce the noise at the cost of signal resolution. Technique to subtract fixed pattern noise from phase and amplitude readouts is developed by \cite{georgiev2015fixed}.

\end{enumerate}
Ideally, the limiting noise sources for the CMOS phase camera have a Gaussian distribution~\cite{toronov2003optimization, rapp2008theoretical}. Under this assumption, the variance of each of the quadrature measurements is constant ($\sigma^{2}$). In this case, the total noise in the amplitude ($\sigma_{A}$) can be estimated as the quadrature sum over individual noise sources. The total phase noise ($\sigma_{\phi}$) is estimated as the ratio of the total noise in amplitude to the mean amplitude of the signal, as phase and amplitude noise should be uncorrelated.

\subsection{Quantitative Noise Measurement and Sensitivity}
\label{subsec:sensitivity}
Here we quantify the temporal and spatial noise of the CMOS phase camera.
The noise performance is characterized using the experimental layout described in Section~\S\ref{sec:exp_sense}. To characterize the DC power saturation levels, we vary the intensity of the 1064~nm laser beam incident on the CMOS sensor. We observe pixel saturation at $6\mu$W of incident power with beam radius of 0.25~mm.
Additional camera performance parameters are summarized in Table~\ref{tab:CameraStats}.

\begin{figure}%
\center
\includegraphics[width=\linewidth]{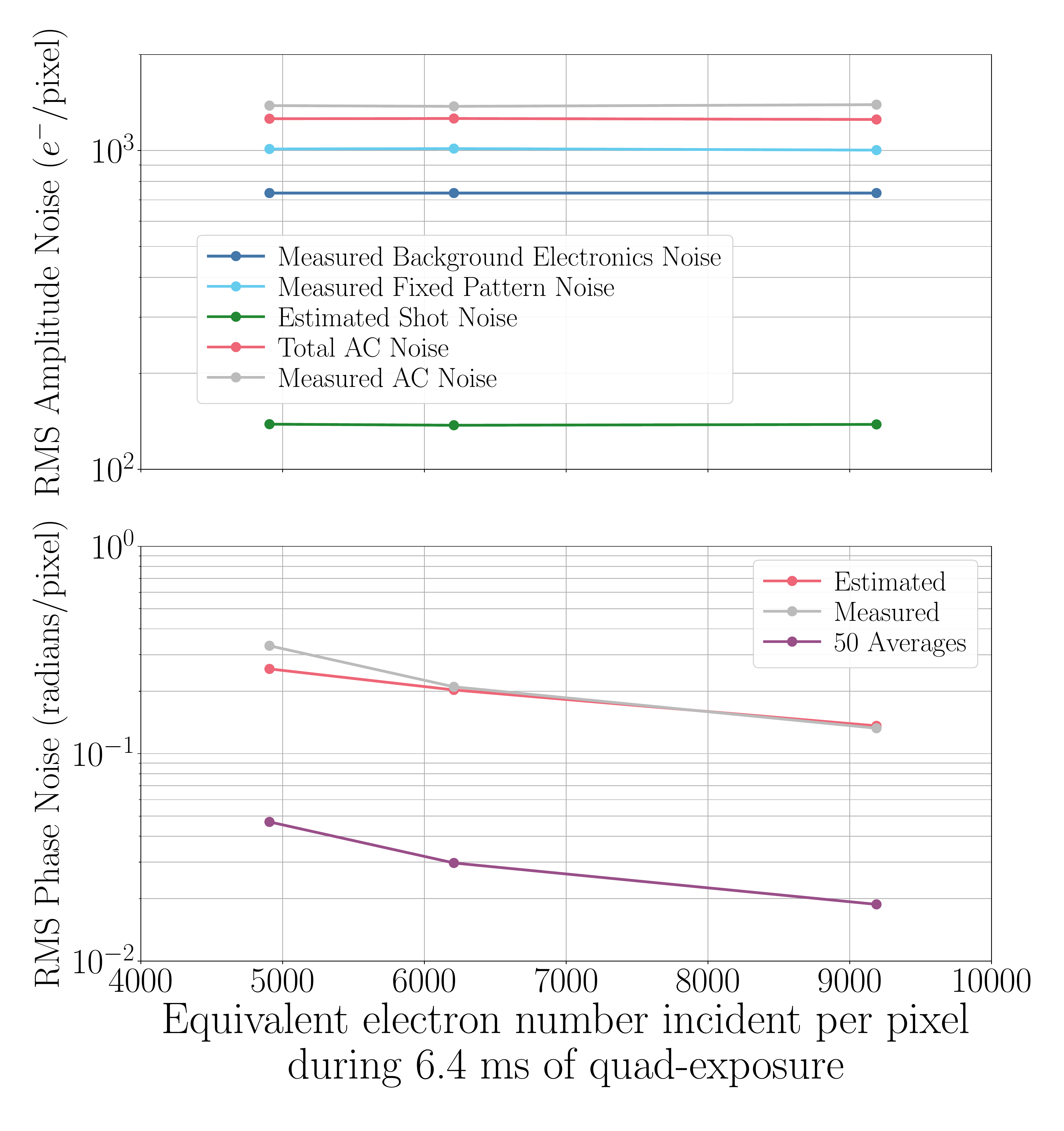}
\caption{
  The figure above shows the phase camera noise as the power modulation index of the incident beam is varied at constant beam intensity. 
  The root-mean-square amplitude noise (top) and phase noise (bottom) per pixel of the camera are calculated from a single image captured (no averaging). 
  As the beam profile is the same across these measurements, the sensor area considered is the same. 
  The shot noise and fixed pattern noise remain constant under these illumination conditions. 
  We find that the total measured per pixel noise in phase and amplitude agrees closely with the sum of the budgeted noise sources. The current prototype of the phase camera is predominantly limited by the electronic noise and the shot noise is a factor of 8 below the total noise. The phase noise improves with higher power modulation index and with averaging of frames. The bottom plot also shows the measured per pixel phase noise with 50 frame averages in purple. 
}
\label{fig:ampnoise}
\end{figure}

To determine the RF sensing capabilities and noise limitations of the phase camera the DC intensity of the beam incident on the sensor is held constant while the power modulation index is varied by sweeping the cavity, as illustrated in Fig.~\ref{fig:exp_setup}. Using this method, we report performance measurements for power modulation index values from zero to 0.046. The AC calibration factor, calculated in section~\S\ref{subsec:calibration}, is used to convert the measured noise into equivalent number of photoelectrons. The total noise in an individual pixel in a single frame is estimated by adding in quadrature photon shot noise, fixed pattern noise, and electronics noise, all of which are measured with an integration time of 32 ms.
The total amplitude and phase noise are measured as the standard deviation of the corresponding image obtained by subtracting two independent illuminated frames and dividing the result by $\sqrt{2}$.  The results in Fig.~\ref{fig:ampnoise} show a close agreement between measured and estimated noise. We find that the shot noise limit is a factor of 8 below the total measured noise and the camera sensitivity is limited by background electronic noise and fixed pattern noise.

The demodulation pixels suppress the DC contribution of the illuminating beam using correlated balance sampling~\cite{LangeThesis, Conde:17}. However, a sufficiently high carrier field intensity will saturate the pixel of the sensor. The dynamic range is the ratio of the saturation point to the noise floor defined as~\cite{LangeThesis}
\begin{equation}
    D/R = 20~\text{log}_{10}\big(\frac{A_{sat}}{\sigma_{dark}}\big),
\end{equation}
where $A_{sat}$ is the pixel saturation value, i.e. maximum digital number and $\sigma_{dark}$ is the dark noise of the pixels. We estimate an operating dynamic range of 75~dB for the phase camera. However, due to the low modulation index of the illuminating beam in our test setup, the camera saturates in DC before reaching the full dynamic range of the amplitude and the phase readouts. 

\begin{figure}
\center
\includegraphics[width=\linewidth]{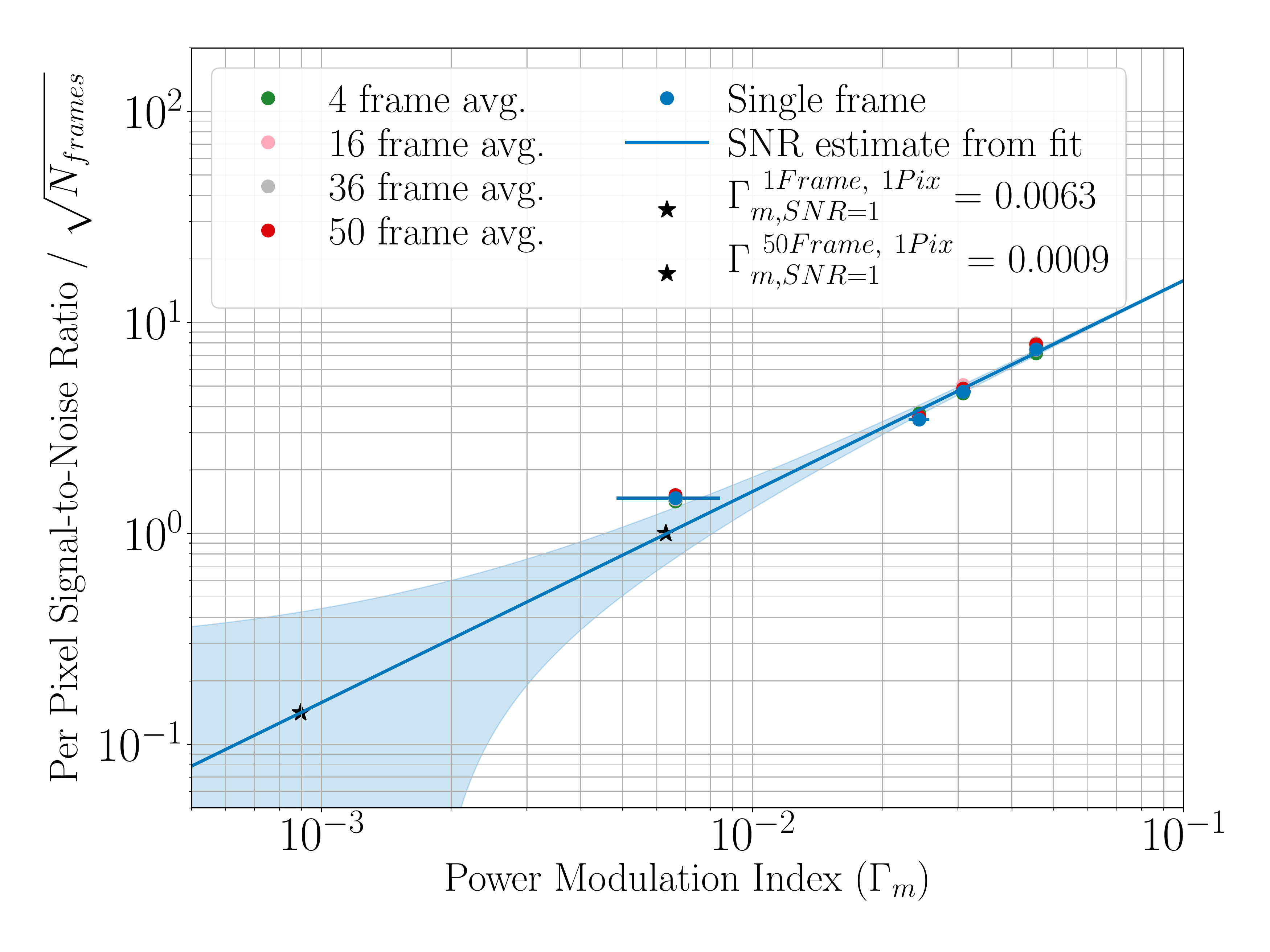}
\includegraphics[width=\linewidth]{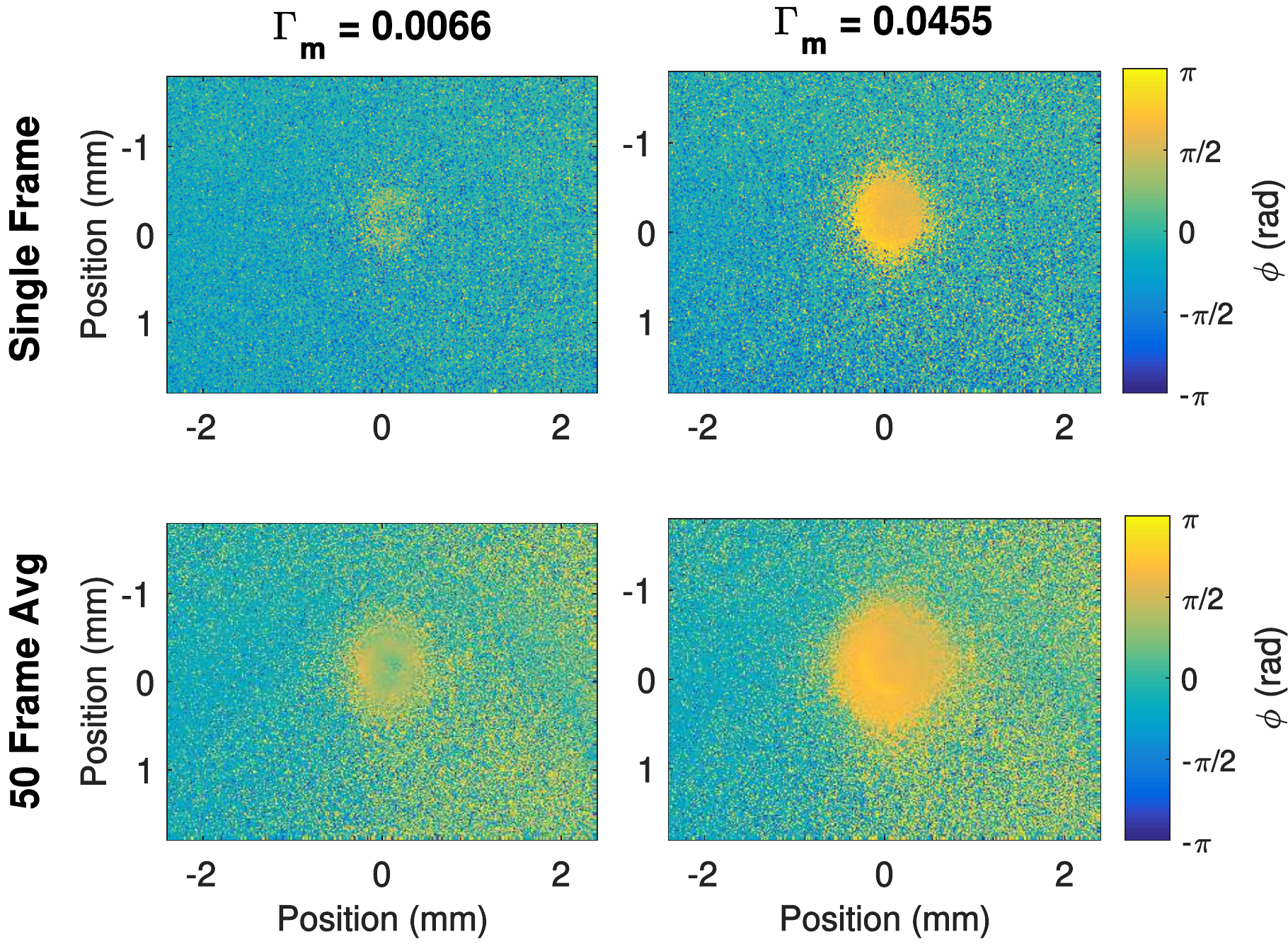}
\caption{Top plot: Single-frame-SNR as a function of $\Gamma_m$. The y-axis shows the single-frame-SNR, defined as SNR/$\sqrt{N_{frames}}$. The horizontal error bars show the experimental errors in the estimation of the $\Gamma_m$ using the calibrated {\it bbPD2} photo detector. We find the SNR improves with the square root of the number of frames and number of pixels, consistent with temporally and spatially independent pixel noise. Pixel averaging can be implemented to improve the SNR at the cost of spatial resolution.
The blue line represents a linear fit through the data. Using the fit, we estimate with 50  averages, the CMOS phase camera is capable of sensing RF signals in each pixel with $\Gamma_m$ as low as 0.0009. Bottom figures: Phase images -- single frame (top row) and with 50  averages (bottom row), for incident beams with low (left) and high (right) power modulation index. The phase resolution improves linearly with SNR.}
\label{fig:detectability}
\end{figure}

Camera performance for each measurement can also be quantified by calculating the signal-to-noise ratio (SNR). We define the SNR as the ratio of amplitude of the demodulated output $A$ (estimated using spatial averaging) to the measured amplitude noise $\sigma_{A}$.
\begin{equation}\label{eqn:SNR}
    SNR = \frac{A}{\sigma_{A}}
\end{equation}
We can define the SNR for individual pixels on a single frame, as well as after spatial and/or temporal averaging, Fig.~\ref{fig:detectability}.
The figure shows the SNR as a function of the power modulation index $\Gamma$ after averaging over a number of frames. The y-axis is scaled by $1/\sqrt{N_{\rm frames}}$, such that the data points line up if the noise in the frames are independent. We find the SNR scales linearly with the modulation index of the illuminating beam at constant beam intensity. 
By fitting a linear model to the data, we determine the phase camera can detect signals above a modulation index of 0.0009 using 50 frame averages without any spatial averaging, see Fig.~\ref{fig:detectability}.

\begin{table*}
  	\begin{center}
  	\begin{tabular}{|p{5.5cm}|ccc|p{6cm}|}
  		\hline \hline 
        \multicolumn{5}{|c|}{\textsc{Capture Settings}} \\
        \hline \hline
        Camera Parameters & ~~Minimum~~ & ~~Typical~~ & ~~Maximum~~ & Comments \\
        \hline
        Frame Rate (fps) &  -  & 7 & 60 (150) & Performance characterization was done at 7 fps. We tested the camera up to 60 fps (live-streaming limit). The camera supports up to 150 fps.  \\
        Quads & 4 & 4 & 6 & Four quad readout is measured with $\pi/2$ phase-stepped demodulation; Six quad readout is measured with $\pi/3$ phase-stepped demodulation. \\
        Sub-frames & 1 & 1 & 4 & The camera allows the capture of 1-4 sub-frames to construct a single frame. Each sub-frame is constructed with the readout of all the quads above. \\
        Quad Integration time (ms) & 1$\%$ $\text{DC}_\text{Exp}$ & 6.4 (15$\%$) & 30$\%$ $\text{DC}_\text{Exp}$ & Dependent on the frame rate, number of sub-quads and sub-frame and the readout time. The quad integration time cannot exceed 30$\%$ of the DC exposure due to the readout and dead time of the CMOS phase camera. \\
        Pixel Resolution ($\mu$m) & & 15 & & This defines the average size of each pixel on the sensor array. \\
        Sensing Area (mm) & & 4.8$\times$3.6 & & The total sensing area of the CMOS sensor. \\
        Demodulation Frequency (MHz) & 5 & 25 & 100 & The performance characterization was done at 25 MHz. The camera supports demodulation frequency up to 100 MHz. The 5 MHz lower limit is due to the RF switch. \\
        \hline \hline 
        \multicolumn{5}{|c|}{\textsc{Illumination Characteristics}} \\
        \hline \hline
        DC Power (nW) [per pixel], estimated with beam size $w_0$ = 0.25~mm & 1.5 & - & 14.6 & The minimum input power for the pixels to sense the input beam and the maximum power before the pixels saturate. \\
        DC Power ($\mu$W) [sensor; for incident beam of size $w_0$ = 0.25~mm] & 0.6 & - & 6.4 & As the spot size of the incident beam increases, both the minimum and maximum DC power can be re-scaled in accordance with per pixel DC power limits above. We caution that this parameter is exposure time dependent. \\
        \hline \hline 
        \multicolumn{5}{|c|}{\textsc{Frame Characterization (at 7~Hz frame rate and 15$\%$ quad integration time)}} \\
        \hline \hline
        Dynamic Range (dB) & & & 75 & The dynamic range of the camera as estimated from the dark noise level of the camera. \\
        $\Gamma_m$: Single Frame & $6.3_{-2.2}^{+2.2}~\cdot 10^{-3}$ & - & 1 & The CMOS phase camera can typically sense as low as 0.0063. \\
        SCR: Single frame & $-50_{-3}^{+3}$ dBc/pixel & - & 0 dBc/pixel & SCR down to -50~dBc/pixel can be resolved in a single frame.\\
        $\Gamma_m$: 50 frames averaged & $0.9_{-0.3}^{+0.3}$ $~\cdot 10^{-3}$  & - & 1 & With 50 frames averaged, the camera can sense incident RF beams as low as 0.0009. \\
        SCR: 50 frames averaged & $-67_{-3}^{+3}$ dBc/pixel & - & 0 dBc/pixel & With 50 frames averaged we can resolve SCR down to -67~dBc/pixel.\\
        Mean AC Noise ($e^-$) & & $10^3$ & & \\
        \hline
  	    \end{tabular}
      \end{center}
    \caption{The table summarizes the range of camera capture setting and the absolute rating for the illumination signals.}
    \label{tab:CameraStats}
\end{table*}

Using equation \ref{eqn:dBc} we can also find the camera's sensitivity limit ($SNR=1$) for resolving the  sideband-to-carrier ratio ${\rm SCR}$ in dBc per frame per pixel:
\begin{equation}\label{eqn:dBc_FRAME_PIXEL}
    {\rm SCR_{LIM}} = 20 \log_{10} \left( \frac{\Gamma_{{\rm SNR}=1}^{1{\rm fr},1{\rm pix}}}{2} \right)  = -50 \, {\rm dBc_{/fr/pix} }.
\end{equation}
Since the frame rate for the test data was $7 ~Hz$, and we only used about $50\%$ of the maximum quad integration time per frame (see Table~\ref{tab:CameraStats}) we find approximately  
\begin{equation}\label{eqn:dBc_SECOND_PIXEL}
    {\rm SCR_{LIM}} = -62 \, {\rm dBc_{/sec/pix} }.
\end{equation}

\section{Signal Modeling for Point Absorbers}
\label{sec:pointAbs}
One of the critical issues limiting the sensitivity of Advanced LIGO is the presence of small (few tens of $um$ diameter) absorptive defects on the test masses, referred to as  {\it point absorbers}. Exposed to the laser field in the interferometer arms, these test mass defects cause local heating and result in local optical path length distortions for the laser field. These optical distortions excite higher-order modes in LIGO's coupled cavities, leading to excess optical loss and limiting the sensitivity of the detectors. Furthermore, the path distortions affect the carrier and sideband phase fronts differently, thus deteriorating the alignment and angular control error signals. A phase camera is capable of mapping these phase front distortions. Error signals can be extracted from the camera output and can be used to control corrective actuators. Unlike conventional quadrant photodiodes, the phase cameras offer a high spatial resolution to resolve the phase front changes due to the point absorbers. 

\begin{figure}
    \centering
    \includegraphics[width=\linewidth]{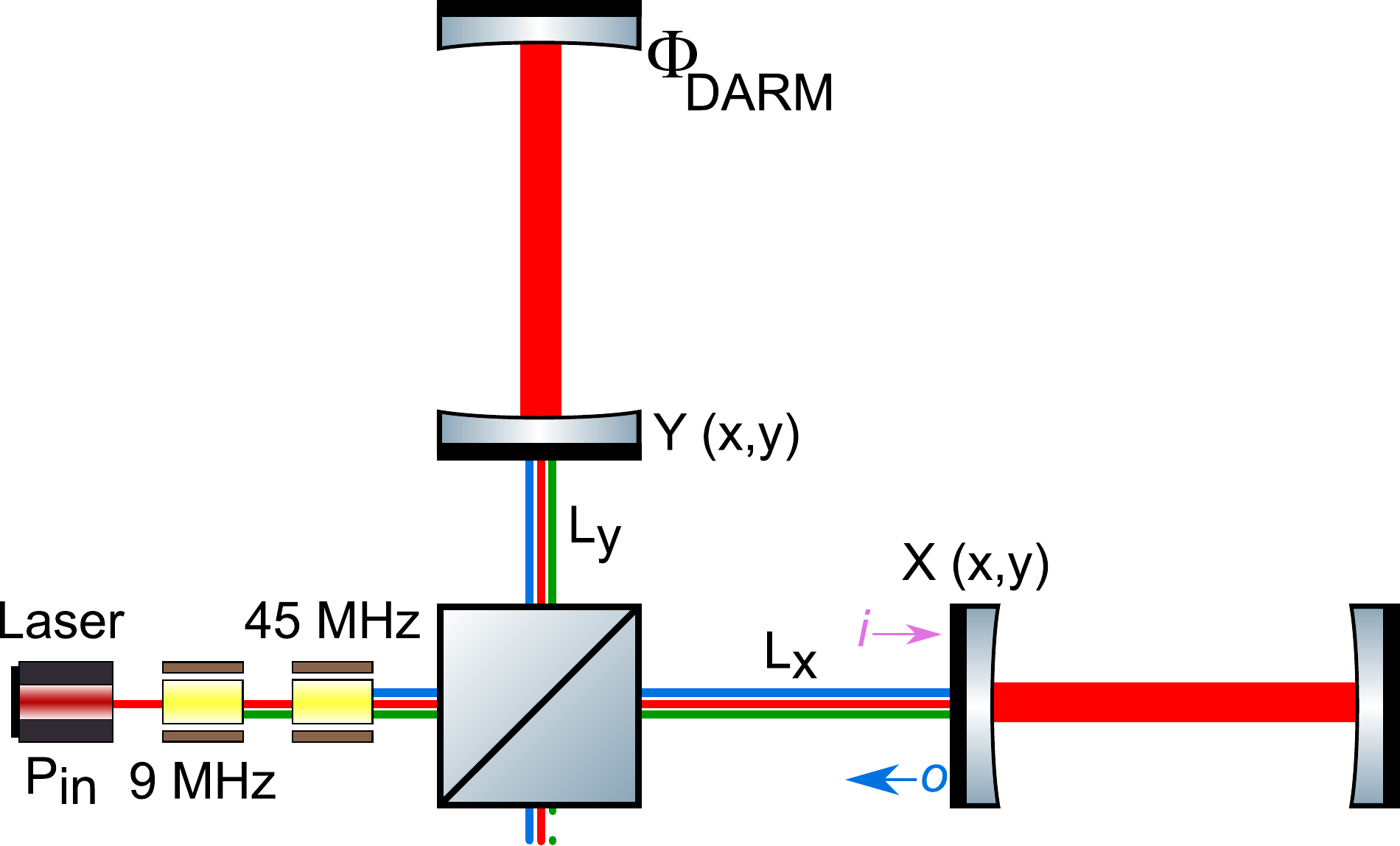}
    \caption{The figure illustrates the simplified Advanced LIGO setup for point absorber modeling. To simplify the simulation we do not include the effects of the power and the signal recycling cavities. Instead, we model the input power to be equal to the power at the beamsplitter during the Advanced LIGO O3 run~\cite{comm_O3}. The other parameters of the interferometer are summarized in the table~\ref{tab:SimParams}. The optical path distortions due to the point absorbers is modeled with a Lorentzian profile for each of the input test masses. The carrier and the sideband fields are calculated under plane beam/paraxial approximation and the corresponding beat signals are calculated at the anti-symmetric port of the beamsplitter. }
    \label{fig:PAModel}
\end{figure}

\begin{table}
  	\begin{center}
  	\begin{tabular}{|p{4cm}c|}
  		\hline \hline 
        \multicolumn{2}{|c|}{\textsc{Simulation Parameters}} \\
        \hline \hline
        Differential arm offset & 10~pm \\
        Arm cavity finesse & 446 \\
        $P_{\text{Homodyne}}$ ref. Beam & 50 mW \\
        $P^{\text{BS}}_{\text{in}}$ & 1500 W \\
        $w^{\text{BS}}_{0}$ & 6~cm \\
        $L_{\text{Schnupp}}$ & 8~cm \\
        Carrier recycling gain  & 40 \\
        $45$~MHz recycling gain  & 2 \\
        $45$~MHz phase mod. index  &  \\
        \,\,\,\,\,\,\,\,at input & 0.18 \\
        \,\,\,\,\,\,\,\,at beamsplitter &0.04 \\
        \hline
  	    \end{tabular}
      \end{center}
    \caption{Simulation Parameters of the interferometer shown in Fig.~\ref{fig:PAModel}. They approximately correspond to the Advanced LIGO parameters \cite{comm_O3}.}
    \label{tab:SimParams}
\end{table}

To get an approximate estimate of the phase camera's ability to sense the effect of point absorbers LIGO's input test masses, we present a simplified model without considering the coupled cavity layout of LIGO, shown in Fig.~\ref{fig:PAModel}. We initially also ignore Gouy phase shifts because beamsplitter and input test masses are essentially in the same Gouy phase. We will get back to the effect of the output beam Gouy phase shift.
While these are oversimplifications, they still allow us the estimate the required sensitivity to pick up the point absorber phase distortions in a phase camera image taken at the interferometer anti-symmetric port.
We consider two scenarios, the current Advanced LIGO, which uses DC readout, and the A+ upgrade, which uses homodyne readout without differential arm DC offset. We choose the model parameters in accordance with the existing Advanced LIGO facilities. We assume the carrier and the 9~MHz and 45~MHz sidebands (with modulation index {$\Gamma$}) are incident at the 50/50 beamsplitter. Considering the approximate power-recycling gain of 40 for the carrier and 2 for the 45MHz sidebands we can estimate the power modulation index of the beam incident on the beamsplitter~\cite{comm_O3, craig2021}. The beam then propagates from the beamsplitter to the input test mass in each of the arm cavities. We model the response of the point absorbers to first-order, which affects the phase of the sidebands, but leaves the carrier unperturbed. The carrier experiences only a phase shift due to the DARM offset between the two arm cavities: 

\begin{align}\label{eqn:iq}
r_x^{car} & = + 1 \cdot e^{i\Phi_{DARM} /2} & r_y^{car} & = + 1 \cdot e^{-i\Phi_{DARM} /2} \\
r_x^{sb} & =  - 1 \cdot e^{i 2 k \text{X} (\Vec{x},\Vec{y})} & r_y^{sb} & =  - 1 \cdot e^{i 2 k \text{Y} (\Vec{x},\Vec{y}) }
\end{align}
\noindent
where $\text{X}(\Vec{x},\Vec{y})$ and $\text{Y}(\Vec{x},\Vec{y})$ are the one-way transmission maps encoding the optical path distortions. Under these simplified assumptions one can calculate the fields of the carrier and sidebands at the anti-symmetric port of the beamsplitter. The beat map between the carrier or the reference beam with the sideband in $\mathcal{I}$ and $\mathcal{Q}$ is given by
\begin{align}
    \mathcal{I}(x,y)&=\mathbb{R}e(sb_{+}^{*} c + c^{*} sb_{-} ) \\
    \mathcal{Q}(x,y)&=\mathbb{I}m(sb_{+}^{*} c + c^{*} sb_{-} )
\end{align}

\begin{figure}
\center
\includegraphics[width=\linewidth]{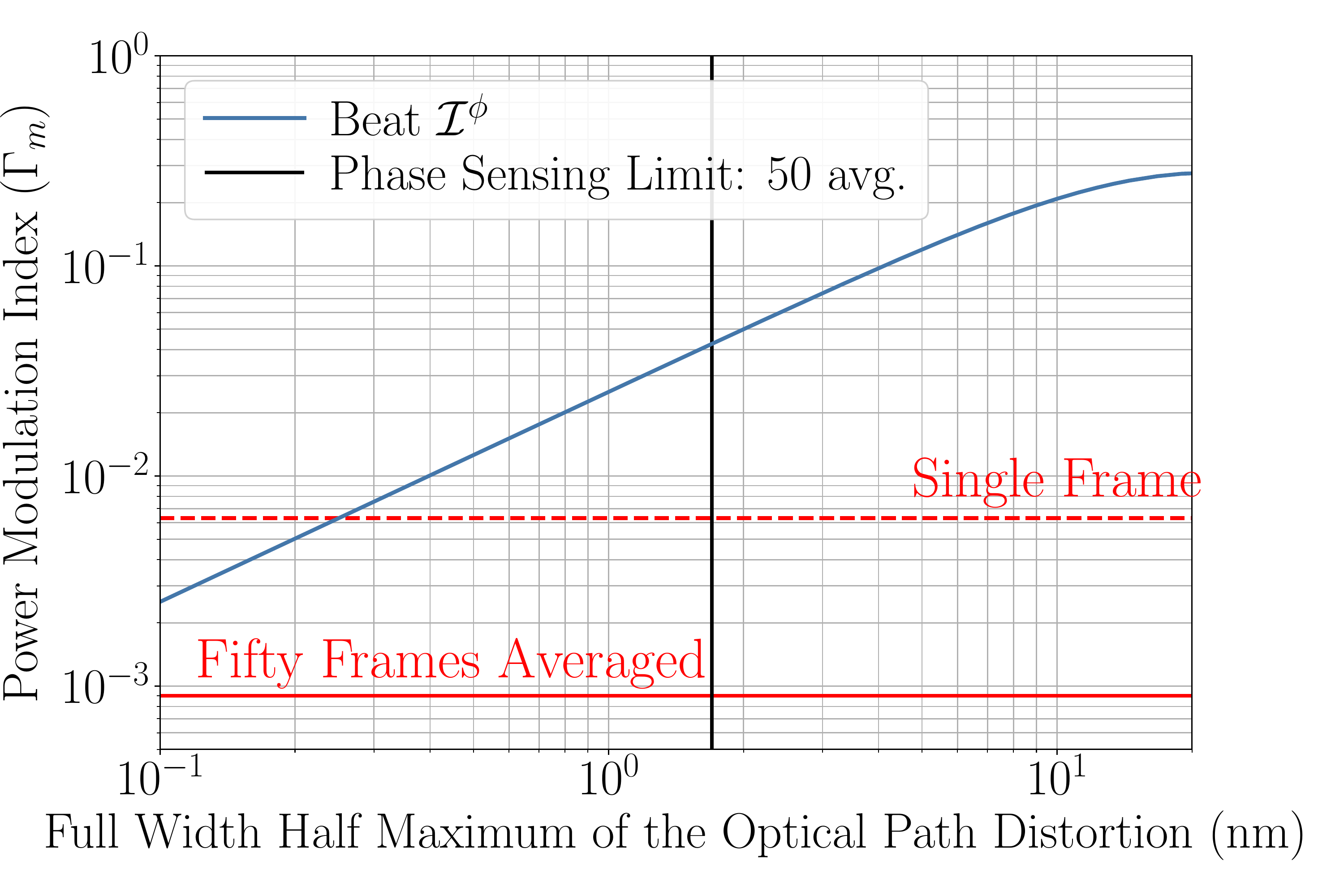}
\caption{Size of the expected power modulation index in the interferometer anti-symmetric port $\mathcal{I}$-quadrature, as a function of the optical path distortions (OPD) due to a point absorber on an input test mass. 
The power modulation index scales linearly with the OPD before it plateaus to a constant, which is caused due to the sidebands, 9~MHz and 45~MHz, leaking through the AS port. 
The red line show the minimum power modulation index that the CMOS phase camera can sense after 50 averaged frames. In homodyne readout, where there is no large orthogonal $\mathcal{Q}$-quadrature signal to compete with, we can sense distortions greater than 0.1~nm using 50 averaged frame.
In the current DC readout there is a large $\mathcal{Q}$-quadrature signal due to the differential arm DC offset. 
The phase resolution of the CMOS phase camera will thus limit sensing capabilities. The dashed black line represents the per pixel phase resolution limit after 50 frame averages (see Fig.~\ref{fig:ampnoise}). 
Assume the camera is place in the appropriate Gouy phase, we can resolve optical path distortions greater than 2~nm in the DC readout scheme of current Advanced LIGO detectors. Typical optical path distortions due to these point absorbers in Advanced LIGO ranges from few nanometers to a few hundreds of nanometer ~\cite{brooks2021point, comm_O3}.
The parameters of the simulation are summarized in Table~\ref{tab:SimParams}.}
\label{fig:PA_res}
\end{figure}

One can analytically show that in the Gouy phase of the beamsplitter ($\phi=0$) the beat in $\mathcal{I}(x,y)$ and $\mathcal{Q}(x,y)$ between the carrier and the sidebands at the AS port is first-order independent of the optical path distortions due to the point absorber. We recover the usual DC readout terms:
\begin{align}\label{eq:IQzero}
    \mathcal{I}(\Vec{x},\Vec{y}) &= 0  \\
    \mathcal{Q}(\Vec{x},\Vec{y}) &= \Gamma p_{in} (\Vec{x},\Vec{y})  \sin(\frac{\Phi_{DC}}{2}) \sin(\frac{\omega_s L_s}{c})
\end{align}
However, this does not remain true at every Gouy phase in the readout beam. The small point absorber distortion evolves differently with Gouy phase. For simplicity, we can assume that the distortion fields $\text{X} (\Vec{x},\Vec{y})$ and $\text{X} (\Vec{x},\Vec{y})$ contains only one higher-order mode of order $N=l+m$. ($l$ and $k$ are for example the Hermite-Gauss mode orders.) Then the $\mathcal{I}$ readout quadrature become
\begin{align}\label{eq:Iphi}
    \mathcal{I}^{\phi} &= k \Gamma p_{in} \sin(\frac{\Phi_{DC}}{2}) \cos(\frac{\omega_s L_s}{c}) \sin(N \phi) (\text{X} - \text{Y}) 
\end{align}
where the spatial shape is given by the beat of the fundamental and the N-th order mode, $\Psi_N(\Vec{x},\Vec{y})\cdot \Psi_0(\Vec{x},\Vec{y})$. 
Equation \ref{eq:Iphi} can be generalized by expanding $\text{X} (\Vec{x},\Vec{y})$ and $\text{X} (\Vec{x},\Vec{y})$ in terms of higher order Gaussian modes, which is straight forward, a little complicated, and not necessary if we are only interested in the camera sensitivity limitation.

The presence of a large beat signal in the $\mathcal{Q}$ quadrature from the DC readout scheme means that we also have to worry about the camera phase resolution. The beat map phase rotation is given by
\begin{align}\label{eq:IpQphi}
    \frac{\mathcal{I}}{\mathcal{Q}}^{\phi} &= k \cot(\frac{\omega_s L_s}{c}) \sin(N \phi) (\text{X} - \text{Y}) 
\end{align}

During the A+ upgrade Advanced LIGO will switch to a homodyne readout scheme, reducing the DC offset to zero, and instead using a separate carrier reference beam as local oscillator. We assume that the reference beam has about the same amplitude as the current carrier due to the DC offset, as this permits using the same sensing and readout electronics. Thus, the expression \ref{eq:Iphi} for sensing the effect from the point absorbers remains essentially the same, with one difference: For diagnostic purposes we now have control over the phase $\alpha_{\text{Hom}}$ of the reference beam. The $\mathcal{I}$ quadrature thus becomes
\begin{align}\label{eq:IphiHom}
    \mathcal{I}^{\phi} &= k \Gamma \sqrt{p_{ref} p_{in} } \cos(\frac{\omega_s L_s}{c}) \sin(N \phi-\alpha_{\text{Hom}}) (\text{X} - \text{Y})
\end{align}
Thus we can pick the beamsplitter Gouy phase $\phi=0$ a $90\deg$ rotated reference beam, $\alpha_{\text{Hom}}=\pi/2$, removing the large beat signal in the $\mathcal{Q}$ quadrature and avoiding the phase resolution limitation of the phase camera.

The power modulation index of the beat signal is given by
\begin{equation}\label{eq:gamma_PA}
    \Gamma_{\mathcal{I}} = \frac{2 \mathcal{I}}{DC};~~~
    \Gamma_{\mathcal{Q}} = \frac{2 \mathcal{Q}}{DC}
\end{equation}
Figure Eq.~\ref{fig:PA_res} shows the size of the expected power modulation index signal as a function of the optical path distortion (OPD). The red horizontal dashed line represents the approximate power modulation index sensitivity limit of the phase camera for each pixel. The vertical black dashed lime corresponds to the phase resolution limit of the camera, and is relevant in the presence of a large signal in the orthogonal quadrature due to the interferometer differential arm fringe offset.
We expect the phase camera to be sensitive enough to pickup optical path distortions greater than about $\sim 2~nm$ where the readout is limited by phase resolution of the phase camera. Otherwise, the phase camera is capable of sensing OPD due to point absorbers as low as $\sim 0.1~nm$. Typically, the OPD caused due to point absorbers ranges between tens up to a few hundreds of nanometers \cite{brooks2021point, comm_O3}. Currently Hartmann wavefront sensors are used to image point absorbers in aLIGO directly. These sensors map the point absorbers onto the surface of the test masses ~\cite{Brooks:07}, but do not measure their impact on the interferometer. The CMOS phase camera does not directly image the point distortions, but instead measures the change in the interferometer phase front at the AS port caused due to these point defects.

\subsection{Implications for Gravitational Wave Detectors}\label{subsec:PAImp}
The simple interferometer model presented above leads us to conclude that the CMOS phase camera developed in our group is capable of diagnosing the effect of point absorbers on the LIGO input test masses when installed at the interferometer anti-symmetric port. The exact Gouy phase of the camera will matter though, as there is no signal in the beamsplitter Gouy phase. Having a separate local oscillator reference beam, either as part of the homodyne readout or as a separate local oscillator for the phase camera, will simplify the image analysis.
Lastly, we note that the model presented above does not include the signal recycling cavity and the power recycling cavity of the Advanced LIGO detector. While the power recycling cavity only filters the beam incident on the beamsplitter, the signal recycling cavity will spatially filter the effect of the point absorber, cleaning up the mode. However, with a signal recycling mirror transmission around $32\%$, the signal recycling cavity has an extremely low finesse, preserving the distortion signal, but also making modeling rather complicated.

\section{Discussion}\label{sec:results}
We demonstrate a CMOS phase camera that is capable of imaging externally modulated RF beat signals incident on the sensor with high spatial resolution. The noise levels of the camera allow sensing of RF beat signals with a power modulation index as low as 0.0009 with 50 frame averages. The phase camera also has the capability to measure the beat signals at different frequencies and is sensitive to very low incident beam power levels. Lastly, the low latency image acquisition, design compactness, and relatively low cost of the phase camera make it suitable for numerous applications in wavefront diagnostics and sensing.

The primary application of a phase camera in gravitational-wave detectors is for diagnostic purpose, imaging any unexpected phase front distortions, such as for example those induced by point absorbers on test masses (see section \ref{sec:pointAbs}). 
However the phase camera can also be useful for controlling alignment and mode-matching in an interferometer. While quadrant-photodiode and bullseye-photodiode-based schemes, with \cite{magana2019sensing} and without \cite{miller2014length} optical mode converters, have the advantage of higher signal-to-noise, they offer only four ``pixels" across the beam. In particular, 
the 320$\times$240 pixel  resolution of the CMOS phase camera provides the sensing capabilities to operate and control interferometers with higher-order Laguerre-Gauss or Hermite-Gauss modes as the operating resonant mode, a scheme that was proposed to reduce the coupling of thermal noise to the gravitational readout~\cite{PhysRevLett.105.231102, carbone2013generation}.
Additionally, it has been shown that these cameras can be operated synchronously~\cite{Shrestha:2016}, which allows for multiple cameras to simultaneously record the beam in separate Gouy phases. 

In summary, we expect that the CMOS phase camera will be an excellent tool for commissioning Advanced LIGO, A+, and future gravitational wave detectors such as Cosmic Explorer~\cite{CE2020LF} or Einstein Telescope~\cite{ET_design}, and might also have control applications.

\section{Acknowledgment}\label{sec:ack}
This research was funded by the National Science Foundation through the awards NSF PHY-1352511 and NSF PHY-1912536. We also would like thank numerous colleagues in the LIGO, Virgo and KAGRA scientific collaborations for many fruitful discussions.
\vspace{.5cm}

\section{Appendix}\label{sec:app}
\subsection{Imaging of Multiple Modulation Frequencies with the CMOS phase camera}\label{appsec:ff_scheme}

\begin{figure}[h]
\center
\includegraphics[width=\linewidth]{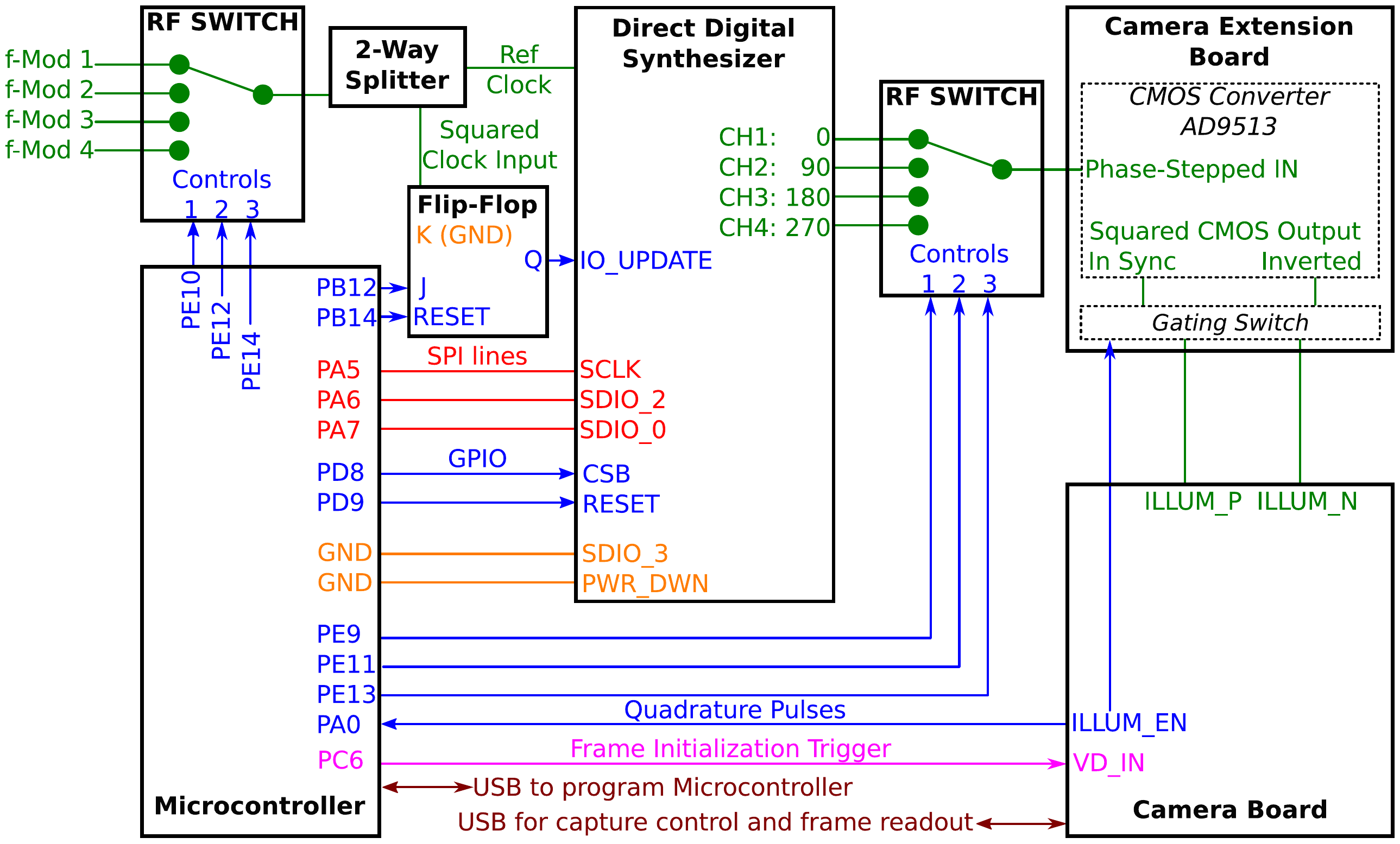}
\caption{Alternate capture scheme mentioned in section~\S\ref{sec:electronics}. It allows simultaneous imaging of RF signals with different modulation frequencies by selecting a separate modulation frequencies for every frame. In this scheme, the microcontroller controls an additional RF switch at the input off the DDS board. A 2-way splitter is used to provide a clock signal input to a flip-flop circuit. The flip-flop ensures a phase-locked output from the DDS by triggering the DDS update on the rising edge of the external local oscillator. The rest of the design block functionality is the same as discussed in Fig.~\ref{fig:signalchain}. The RF switch at the output of the DDS can be discarded if six sub-quads are desired for image capture. In this scenario, the DDS is configured to phase-step one particular channel in synchronization with the quadrature pulses received by the microcontroller.
}
\label{fig:ff_design}
\end{figure}

As discussed in section~\S\ref{sec:electronics}, a flip-flop~(FF) can be integrated with the CMOS phase camera to support four or six quad exposures. The flip-flop also allows the camera to subsequently image beat signals at different modulation frequencies. The electronic setup for this operation is shown in figure~\ref{fig:ff_design}. An RF-switch is used to switch between the modulating frequencies of the local oscillators following the frame capture sequence governed by the microcontroller. The frequency and phase stepped output of the DDS are phase-locked by ensuring the updates are triggered on the rising edge of the local-oscillator by the flip-flop. The functionality of other elements in this scheme is the same as discussed in section~\S\ref{sec:electronics}. This setup is particularly useful for application in Advanced LIGO, which allows the CMOS phase camera to image beat signals at 9~MHz and 45~MHz in real time. Using this technique, the CMOS phase camera can image beat signals at different modulation frequencies with a low-latency of 1~Hz.

\bibliographystyle{apsrev4-1}
\bibliography{PhCam}
\end{document}